\documentclass[11pt]{article}

\usepackage{graphicx}
\usepackage{amsmath}
\usepackage{amssymb}
\usepackage{hyperref}
\usepackage{cite}
\usepackage{times}
\usepackage{fullpage}
\usepackage{ifthen}
\usepackage[T1]{fontenc}

\newcommand{\spacing}[1]{\renewcommand{\baselinestretch}{#1}\large\normalsize}
\spacing{1}

\def\@maketitle{%
  \newpage\spacing{1}\setlength{\parskip}{12pt}%
    {\Large\bfseries\noindent\sloppy \textsf{\@title} \par}%
    {\noindent\sloppy \@author}%
}

\def\arcsec{\hbox{$^{\prime\prime}$}}
\def\arcmin{\hbox{$^{\prime}$}}
\def\degr{\hbox{$^{\circ}$}}
\newcommand{\angstrom}{\text{\normalfont\AA}}

\newcommand{\mnras}{{Mon. Not. R. Astron. Soc.}}
\newcommand{\apj}{{Astrophys. J.}}

\newcommand{\aap}{{Astron. Astrophys.}}
\newcommand{\apjs}{{Astrophys. J. Suppl. Ser.}}

\newcommand{\RNum}[1]{\uppercase\expandafter{\romannumeral #1\relax}}

\newenvironment{addendum}{%
    \setlength{\parindent}{0in}%
    \small%
    \begin{list}{Acknowledgements}{%
        \setlength{\leftmargin}{0in}%
        \setlength{\listparindent}{0in}%
        \setlength{\labelsep}{0em}%
        \setlength{\labelwidth}{0in}%
        \setlength{\itemsep}{12pt}%
        }
    }
    {\end{list}\normalsize}

\newcommand*{\addendumlabel}[1]{\textbf{#1}\hspace{1em}}

\newenvironment{methods}{%
    \section*{Methods}%
    \setlength{\parskip}{6pt}%
    }{}

\title{Blue large-amplitude pulsators\\as a new class of variable stars}

\date{}

\begin{document}

\maketitle

\newenvironment{affiliations}{%
    \setcounter{enumi}{1}%
    \setlength{\parindent}{0in}%
    \slshape\sloppy%
    \begin{list}{\upshape$^{\arabic{enumi}}$}{%
        \usecounter{enumi}%
        \setlength{\leftmargin}{0in}%
        \setlength{\topsep}{0in}%
        \setlength{\labelsep}{0in}%
        \setlength{\labelwidth}{0in}%
        \setlength{\listparindent}{0in}%
        \setlength{\itemsep}{0ex}%
        \setlength{\parsep}{0in}%
        }
    }{\end{list}\par\vspace{12pt}}

\renewenvironment{abstract}{%
    \setlength{\parindent}{0in}%
    \setlength{\parskip}{0in}%
    \bfseries%
    }{\par\vspace{0pt}}


{\bf
{\noindent \author{Pawe{\l} Pietrukowicz$^{1*}$, Wojciech A. Dziembowski$^{1,2}$,
Marilyn Latour$^{3}$, Rodolfo Angeloni$^{4,5,6}$, Rados{\l}aw Poleski$^{1,7}$,
Francesco di Mille$^{8}$, Igor Soszy\'nski$^{1}$, Andrzej Udalski$^{1}$,
Micha{\l} K. Szyma\'nski$^{1}$, {\L}ukasz Wyrzykowski$^{1}$, Szymon Koz{\l}owski$^{1}$,
Jan Skowron$^{1}$, Dorota Skowron$^{1}$, Przemek Mr\'oz$^{1}$,
Micha{\l} Pawlak$^{1}$ \& Krzysztof Ulaczyk$^{1,9}$}
}
}


\begin{affiliations}
\item Warsaw University Observatory, Al. Ujazdowskie 4, 00-478 Warszawa, Poland
\item Nicolaus Copernicus Astronomical Center, ul. Bartycka 18, 00-716 Warszawa, Poland
\item Dr. Karl Remeis-Observatory \& ECAP, Astronomical Institute, Friedrich-Alexander University Erlangen-Nuremberg, Sternwartstr. 7, 96049, Bamberg, Germany
\item Departamento de F\'isica y Astronom\'ia, Universidad de La Serena, Av. Cisternas 1200 Norte, La Serena, Chile
\item Instituto de Investigaci\'on Multidisciplinar en Ciencia y Tecnolog\'ia, Universidad de La Serena, Av. Ra\'ul Bitr\'an 1305, La Serena, Chile
\item Gemini Observatory, Casilla 603, La Serena, Chile
\item Department of Astronomy, Ohio State University, 140 W. 18th Ave., Columbus, OH 43210, USA
\item Las Campanas Observatory, Casilla 601, La Serena, Chile
\item Department of Physics, University of Warwick, Gibbet Hill Road, Coventry, CV4 7AL, UK \\
$^{*}$e-mail: pietruk@astrouw.edu.pl
\end{affiliations}

\begin{abstract}
Regular intrinsic brightness variations observed in many stars are
caused by pulsations. These pulsations provide information on the global
and structural parameters of the star. The pulsation periods range from
seconds to years, depending on the compactness of the star and
properties of the matter that forms its outer layers. Here, we report
the discovery of more than a dozen of previously unknown short-period
variable stars: blue large-amplitude pulsators. These objects show
very regular brightness variations with periods in the range of
20--40~min and amplitudes of 0.2--0.4~mag in the optical passbands.
The phased light curves have a characteristic sawtooth shape, similar
to the shape of classical Cepheids and RR Lyrae-type stars pulsating
in the fundamental mode. The objects are significantly bluer
than main sequence stars observed in the same fields, which
indicates that all of them are hot stars. Follow-up spectroscopy
confirms a high surface temperature of about 30,000~K. Temperature
and colour changes over the cycle prove the pulsational nature
of the variables. However, large-amplitude pulsations at such short
periods are not observed in any known type of stars, including hot
objects. Long-term photometric observations show that the variable stars
are very stable over time. Derived rates of period change are of the order
of $10^{-7}$ per year and, in most cases, they are positive. According
to pulsation theory, such large-amplitude oscillations may occur in
evolved low-mass stars that have inflated helium-enriched envelopes.
The evolutionary path that could lead to such stellar configurations
remains unknown.
\end{abstract}

Thousands of pulsating stars have been discovered in the Milky Way
and other galaxies of the Local Group over the last decades, mainly thanks
to large-scale variability surveys. The OGLE survey \cite{2015AcA....65....1U}
has created the largest ever collection of variable stars
\cite{2013AcA....63...21S,2014AcA....64..177S,2015AcA....65..297S,2016AcA....66..405S,2015AcA....65..313M}.
In 2016, this collection reached almost a million objects, nearly half of
which are pulsating stars. Among the detected pulsators are long-period
variables, $\delta$~Cephei, RR Lyrae and $\delta$~Scuti-type stars.
By monitoring about a billion stars in the sky, we are able to find extremely
rare objects, such as the blue large-amplitude pulsators (BLAPs) presented here.

The first unusual ultra-short-period variable with a sawtooth
light curve was found in the OGLE Galactic Disk field towards the
constellation of Carina during our search \cite{2013AcA....63..379P}
for variable objects with periods shorter than 1~h, conducted in 2013.
Owing to the location in an obscured area of the disk at unknown distance,
it was not possible to obtain information on the luminosity of this object.
With a period of about 28.26~min and a light curve resembling
fundamental-mode pulsators, the star was tentatively classified as a
$\delta$~Sct-type variable and named OGLE-GD-DSCT-0058. However, its
amplitude of about 0.24~mag in the $I$ band was higher than had ever been
observed among the shortest-period $\delta$~Sct variables. It is several
times higher than in a dozen known $\delta$~Sct pulsators that have
a dominant period below 40~minutes \cite{2000A&AS..144..469R}.
Without information on the colour variations, it was not clear whether
the observed variability was due to pulsations. For these reasons,
we placed this object on the list of puzzling OGLE variables for a
low-resolution spectroscopic follow-up in 2014. The spectrum obtained
\cite{2015AcA....65...63P} was characterized by a continuum rising
strongly to the blue, with superimposed hydrogen and helium lines
indicating high effective temperature $T_{\rm eff}$. It looked
similar to those of moderately helium-enriched hot subdwarfs
of O and B type (sdOB). By fitting a model atmosphere appropriate for
typical hot subdwarfs, we found a temperature around 33,000~K,
corresponding to spectral type O9, and surface gravity of
${\rm log}~g\approx5.3$ dex. This temperature is far too high for
$\delta$~Sct-type stars, which have $T_{\rm eff}$ between 6,000 and 10,000~K
or spectral types between A0 and F9. The high surface gravity ruled
out the possibility that it could be a pulsating main-sequence star of
$\beta$~Cephei type. Moreover, pulsators of this type have periods
at least three times as long and typically much lower amplitudes
\cite{2005ApJS..158..193S,2008A&A...477..907P,2008A&A...477..917P}.
This is even more true for another hot main-sequence pulsators,
slowly pulsating B (SPB) stars \cite{2007CoAst.150..167D}. The high
amplitude excluded the possibility that the star was an oscillating
hot subdwarf. Typical amplitudes of the dominant modes
in hot subdwarfs are of millimagnitudes or lower \cite{2016PASP..128h2001H}.
Our object varies much more rapidly and with much higher amplitude than
two known radially pulsating extreme helium stars, BX Cir and V652 Her,
which have periods of 2.6~h and $V$-band amplitudes of about 0.1~mag
\cite{2002A&A...395..535W,2015MNRAS.447.2836J}. These stars are low-mass
supergiants of spectral type B with atmospheres exceptionally poor
in hydrogen ($<0.1$\%). The observed properties of OGLE-GD-DSCT-0058
do not fit any known type of variable star
\cite{2004ESASP.559....1C,2008JPhCS.118a2010E,2015pust.book.....C}.

\begin{figure*}[h!]
\begin{center}
\includegraphics[width=0.65\textwidth]{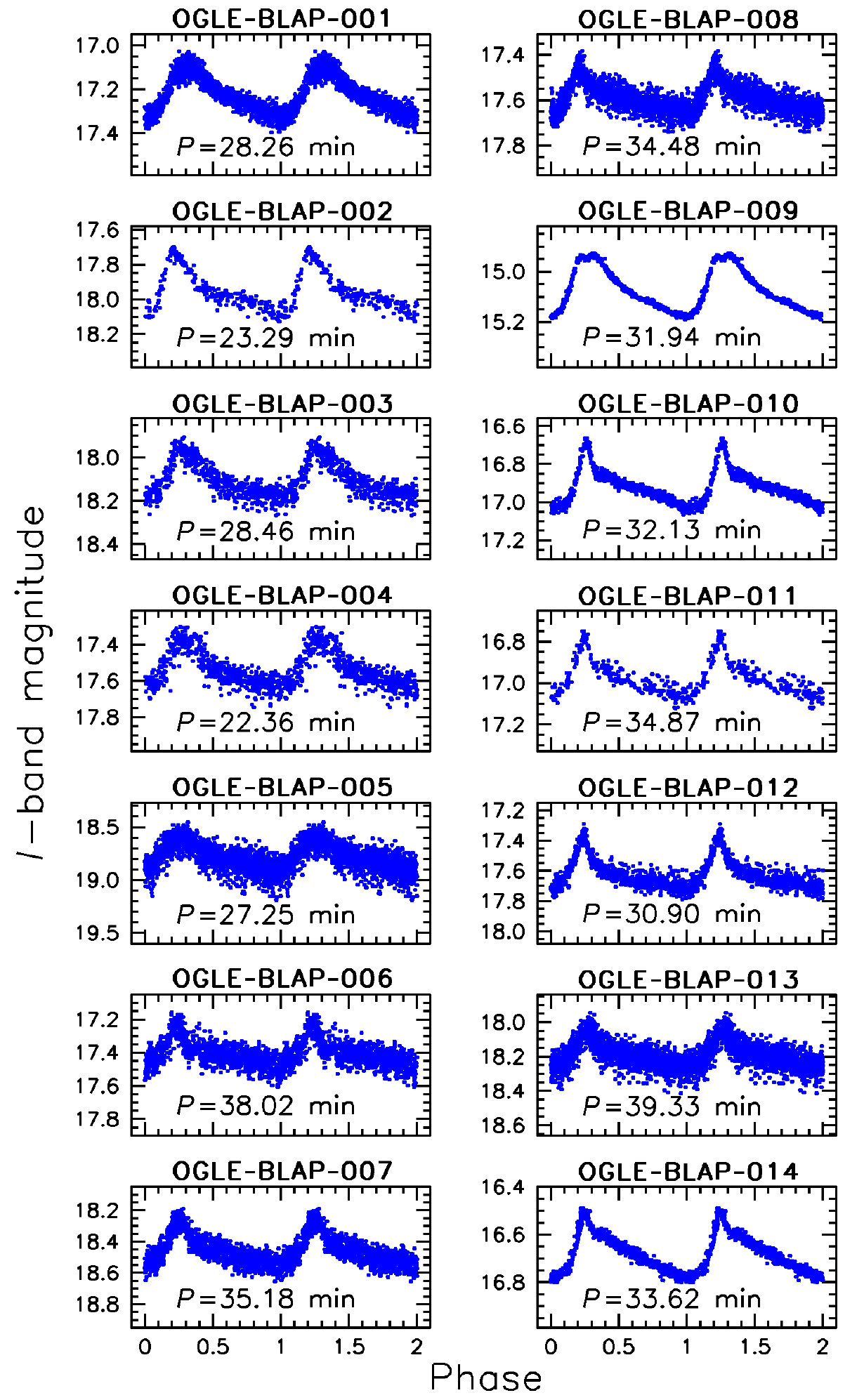}
\end{center}
\caption{\textbf {Phased $I$-band light curves of blue large-amplitude
pulsators (BLAPs) detected by the OGLE survey.} Period, $P$, is given for each
object. The light curve shapes of the new variables are remarkably similar
to the shapes of $\delta$~Sct, RR Lyrae and $\delta$~Cep-type stars
pulsating in the fundamental mode, but those have much longer periods.}
\label{fig:curves}
\end{figure*}

\begin{figure*}[h!]
\begin{center}
\includegraphics[width=0.65\textwidth]{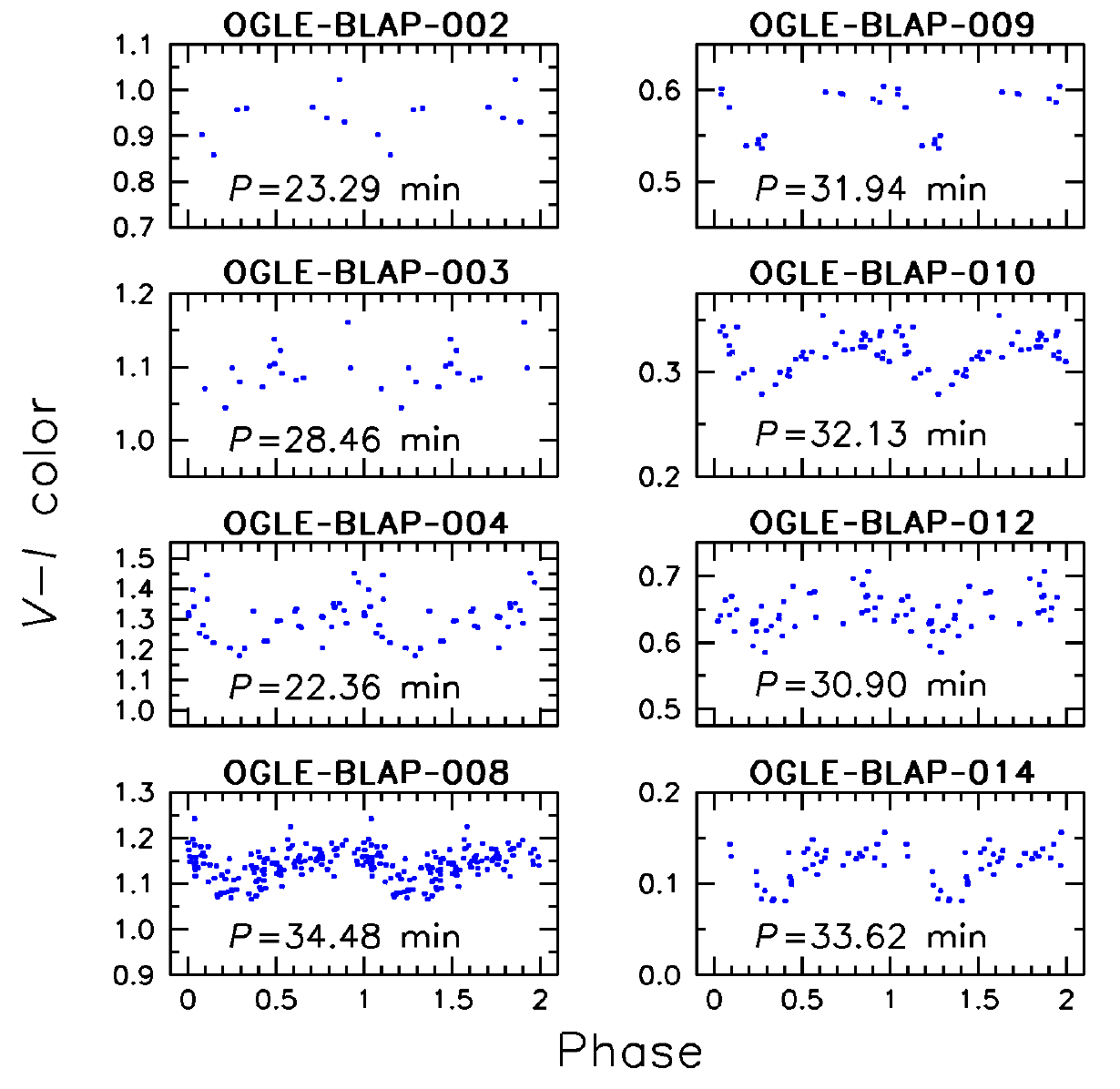}
\end{center}
\caption{\textbf {Variations of the $V-I$ colour over the cycle in the
new variable stars.} This certifies that the observed variability is
driven by physical changes in the star, that is, by pulsations.}
\label{fig:color}
\end{figure*}

Our recent search for short-period variables conducted in OGLE fields
in the Galactic Bulge has brought about the discovery of another 13
objects with almost identical photometric behaviour (Fig.~\ref{fig:curves}).
The periods of these variables range from about 22 to 39~min, and their
$I$-band amplitudes are between 0.19 and 0.36~mag. Such high amplitudes
at such short periods are not observed in any known variables. All the light
curves are very similar in shape: they show a fast increase and a
slow decline which is very characteristic for fundamental-mode pulsators.
In none of the new variables is amplitude modulation observed.
Photometric scatter in some light curves results from a high number of
observations and the presence of close neighbouring stars of constant
brightness in the images. All the variables exhibit only one period;
we do not see any signs of binarity.
The pulsational nature of the new variables can now be demonstrated by
evident colour changes over the cycle (Fig.~\ref{fig:color}). The measured
$V-I$ colour index variations are of 0.04 to 0.17~mag. The drop in the
colour index covers approximately a quarter of the cycle and
correlates with the brightness increase. Lower colour index results
from higher effective temperature. This is observed in
radially pulsating stars in which brightness variations are a
consequence of regular temperature and radius variations.

\begin{figure*}[h!]
\begin{center}
\includegraphics[width=0.65\textwidth]{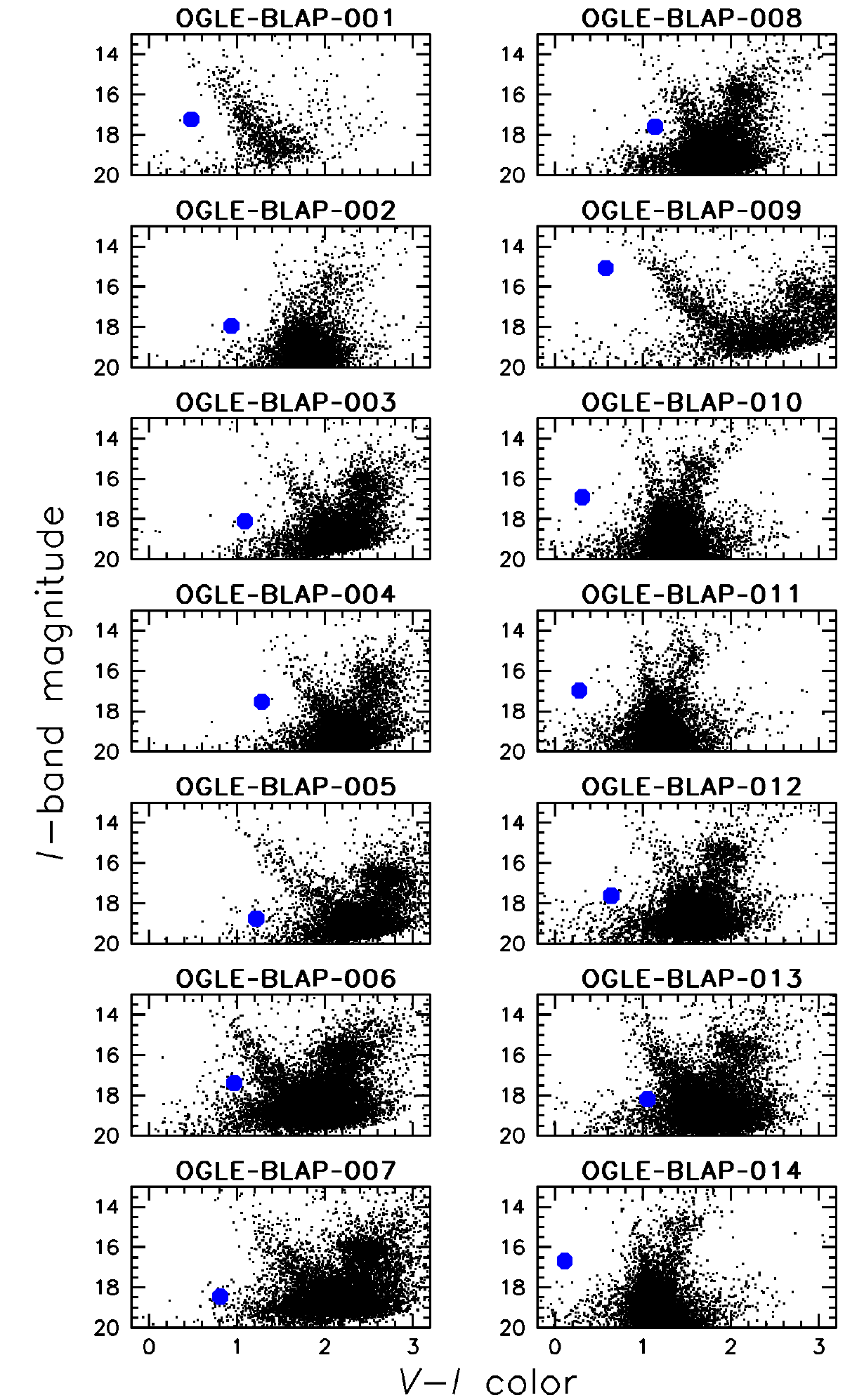}
\end{center}
\caption{\textbf {Colour-magnitude diagrams for stars in the fields
with detected BLAPs.} Large circles mark mean positions of the variables.
Observed brightness and colour variations are roughly of the
size of the blue circles. The new pulsators have significantly bluer
colour than main-sequence stars of the same brightness,
which form the left branch of the V-shaped structure.}
\label{fig:cmd}
\end{figure*}

All 14 detected variables form a homogeneous class of objects
because of their extremely blue colour. Their position in the
colour-magnitude diagrams for stars observed in the same fields clearly
indicate high effective temperatures of the variables (Fig.~\ref{fig:cmd}).
They are all located far blueward of the main-sequence stars of the
same brightness. Based on the photometric properties of the new
objects, which pulsate with exceptionally high amplitudes not previously
observed earlier in any type of hot stars, we propose to call the whole
class of variables 'blue large-amplitude pulsators' (BLAPs). We arranged
the detected variables according to increasing right ascension and
named them OGLE-BLAP-NNN, where NNN is a three-digit consecutive
number. We rename the variable discovered in 2013 as OGLE-BLAP-001,
and henceforth we propose to treat it as the prototype object of the
whole class introduced here.

\begin{figure*}[h!]
\begin{center}
\includegraphics[width=1.0\textwidth]{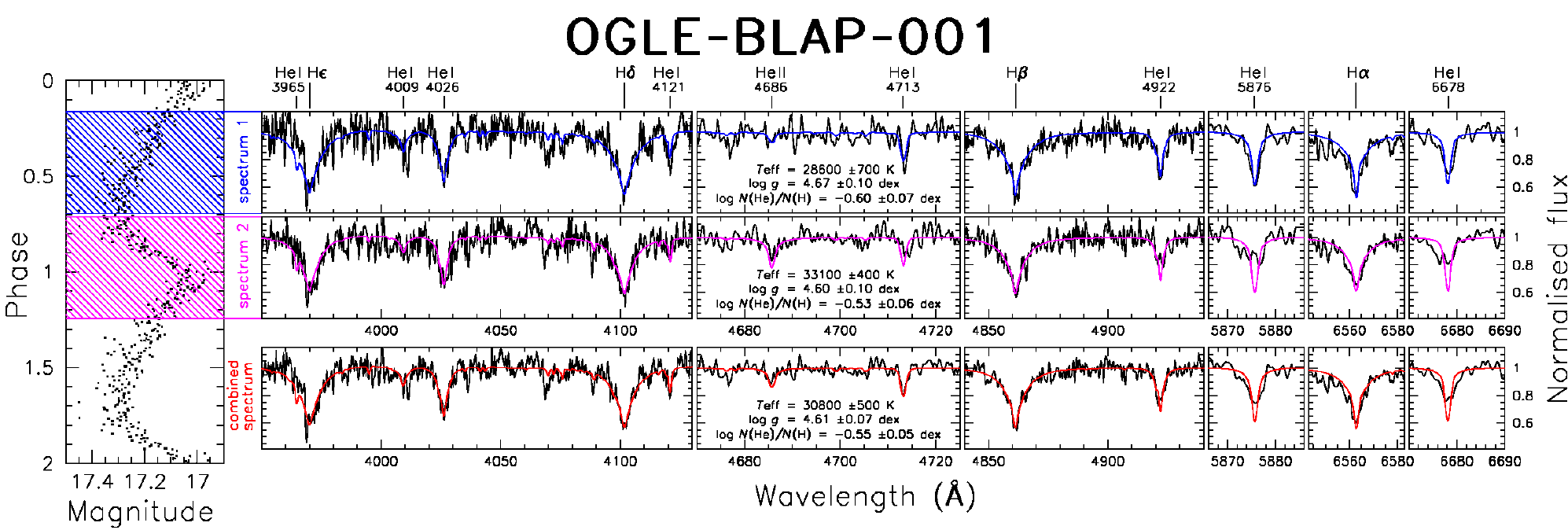}
\end{center}
\caption{\textbf {Magellan-Baade moderate-resolution spectra at opposite
phases of the cycle of the prototype object OGLE-BLAP-001.} Best fits of
stellar atmosphere models to the observed spectra are shown by colour
lines. The different effective temperatures derived indicate that the observed
variability is indeed caused by pulsations. With increasing temperature,
lines of ionized elements become stronger, for example He~{\small II}~4,686~\angstrom.
Short-period pulsations result in striking changes in the shape of neutral
helium lines (He~{\small I}). Parameters derived from the combined
spectrum can be treated as mean values for this prototype object.
The uncertainties shown are formal errors returned by the $\chi^2$
fitting procedure.}
\label{fig:spec1}
\end{figure*}

In 2016, we obtained a pair of moderate-resolution spectra of
OGLE-BLAP-001, exposed one after another, each covering about 53\%
of the period. The spectra were taken around anti-phases of the
cycle to investigate parameter changes. Based on follow-up photometry
for this object, also obtained in 2016, we were able to verify phase coverage
of the new spectra (Fig.~\ref{fig:spec1}). The first spectrum covered most
of the fading branch down to the minimum brightness, and the second one
covered the whole rising branch and a short part of the fading branch.
Our new spectra confirm that there are temperature variations
driven by pulsations. Neutral helium lines clearly change their depth
and shape; the ionized helium line He~{\small II} 4,686~\angstrom~appears
stronger in the second spectrum when the object was hotter. The effective
temperature---derived from fitting model atmospheres---varied from
28,600 to 33,100~K: that is, by about 13\%. Owing to rapid variations
of the physical parameters of the star, the true amplitude of the
temperature changes is likely to be higher. As observed in pulsating stars,
higher temperature corresponds to the maximum brightness. The new
higher-resolution spectra helped us to achieve a better estimate of the
surface gravity in OGLE-BLAP-001. The obtained value of ${\rm log}~g\approx4.6$~dex
is different from what is observed in upper-main-sequence stars
(${\rm log}~g<4.3$~dex \cite{1999AcA....49..119P}) and known sdOB stars
(5.3--6.2~dex \cite{2003A&A...400..939E,2016A&A...589A...1R}).
The atmosphere of OGLE-BLAP-001 seems to be moderately rich in helium,
with a helium-to-hydrogen number ratio of about 0.28.

\begin{figure*}[h!]
\begin{center}
\includegraphics[width=1.0\textwidth]{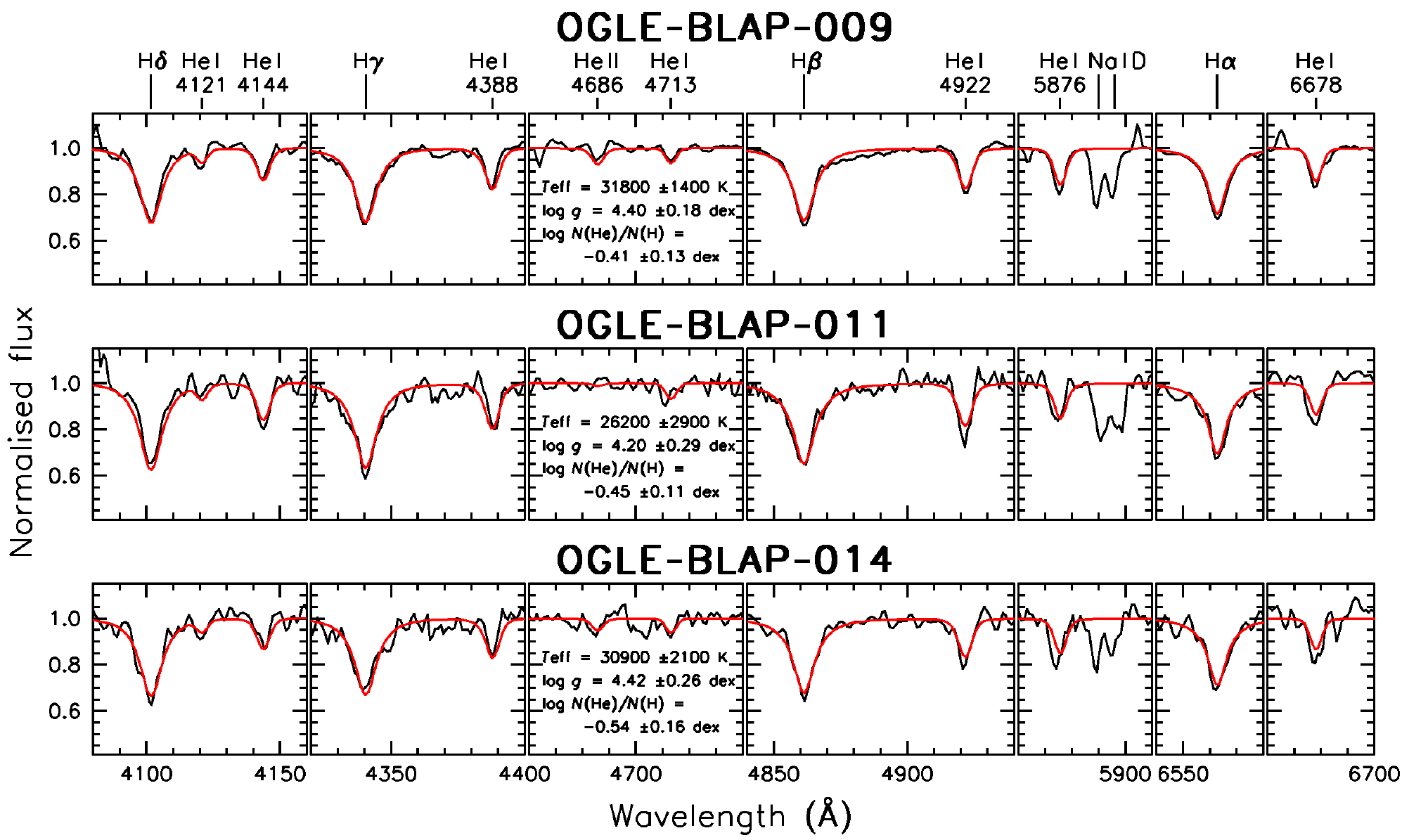}
\end{center}
\caption{\textbf {Gemini-South low-resolution spectra for three BLAPs.}
In the segments, small ticks are every 10~\angstrom.
Best fits of stellar atmosphere models are shown with red lines. The sodium
doublet (Na~{\small I}~D) is of interstellar origin. The uncertainties
are formal errors returned by the $\chi^2$ fitting procedure. Effective
temperatures, surface gravities, and helium abundances derived for these
stars are similar to the values obtained from moderate-resolution spectra
for the prototype object OGLE-BLAP-001. This shows that all the newly
discovered variables form a homogeneous class of objects.}
\label{fig:spec3}
\end{figure*}

In 2016, we also obtained low-resolution spectra of three other stars
of the new class (Fig.~\ref{fig:spec3}): OGLE-BLAP-009, OGLE-BLAP-011
and OGLE-BLAP-014. Within the parameter uncertainties, the atmospheric
properties of these stars are very similar to those of the prototype object.
This shows that pulsations in BLAPs are excited under specific conditions.

The key to understanding the nature of the new pulsators is rates
of period change rates determined from the long-term OGLE observations
(Table~1). We estimate period changes for 11 objects.
The rates are of the order of $10^{-7}$~yr$^{-1}$. Among six
variables with significant period changes ($>3\sigma$), the rates
are negative in two stars, whereas they are positive in four objects.
In three other stars, the changes are very likely positive
(with standard deviation between 2 and 3). Our observations show
the long-term stability of the pulsation period in BLAPs.

\begin{figure*}[h!]
\begin{center}
\includegraphics[width=1.0\textwidth]{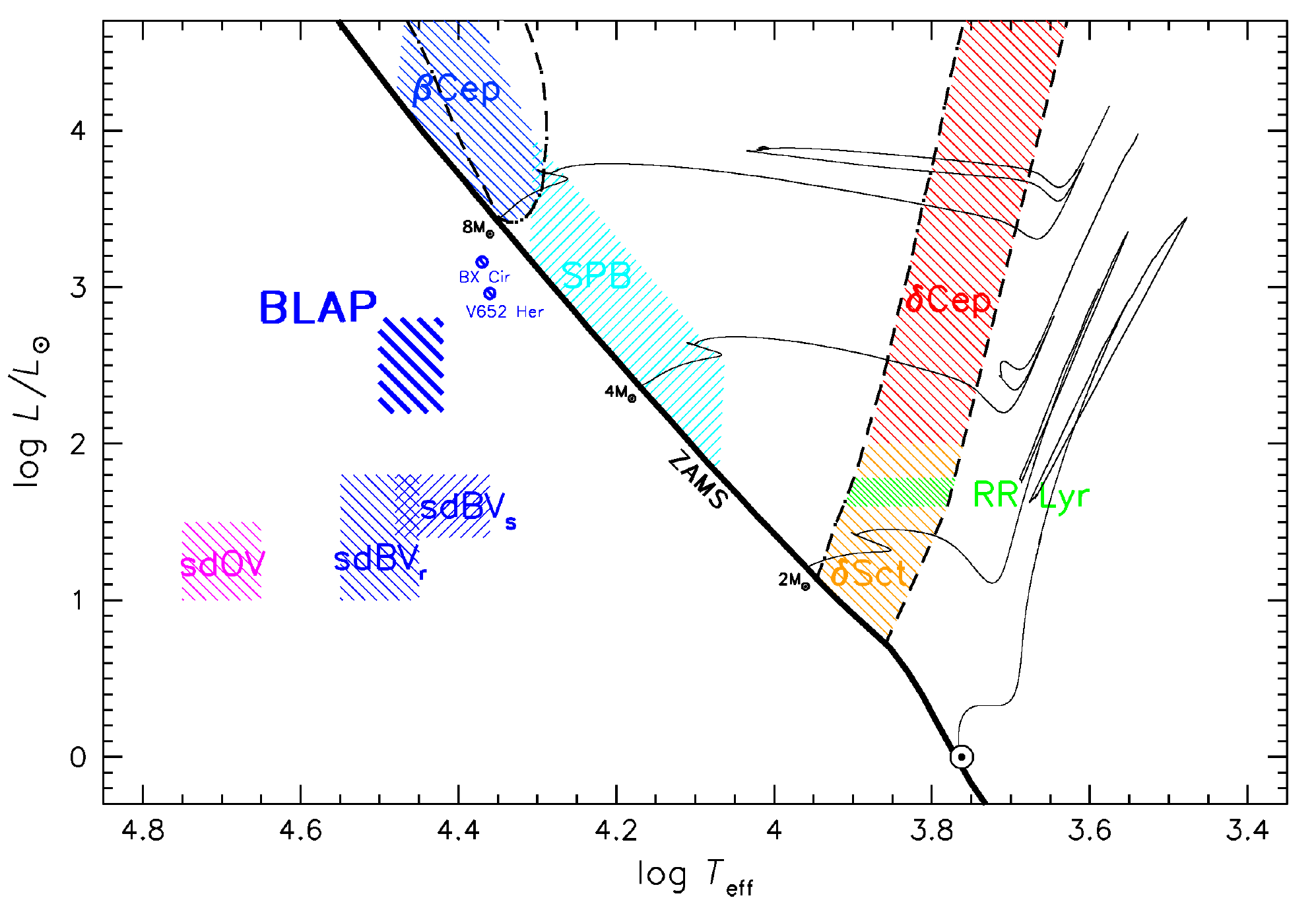}
\end{center}
\caption{\textbf {Location of the BLAPs in the Hertzsprung-Russell
diagram.} The zero-age main sequence (thick solid line),
evolutionary tracks for 1, 2, 4 and 8~M$_{\odot}$ (thin solid lines,
\cite{2006ApJ...642..797P}), edges of the classical instability strip and
instability domain for radial pulsations in the upper main sequence
(dashed lines, \cite{1999AcA....49..119P}) are presented for the metallicity
of $Z=0.02$. Hatched regions show locations of various types of known
pulsating stars in which pulsations are driven by the presence
of He {\small II} bump ($\delta$~Cep, $\delta$~Sct, and RR Lyrae-type
stars) and Z bump in the opacity ($\beta$~Cep-type, SPB stars,
sdO, slowly (s) and rapidly (r) pulsating sdB variables). Different
shading represents different types of modes: backslash (\textbackslash) shading is for
pressure modes whereas slash (/) shading is for gravity modes. Positions
of two known radially pulsating extreme helium stars (BX Cir and
V652 Her) are also marked (small circles). BLAPs are located
in a region not occupied by any known pulsating variables.}
\label{fig:hr}
\end{figure*}

The upper limit on absolute systematic rates of period change implies
that we are dealing with objects slowly evolving on the nuclear timescale.
As the main-sequence phase has been already excluded, one of the
two remaining possibilities is the core helium-burning phase, the same
as for classical Cepheids, RR Lyrae-type stars and sdOB stars. BLAPs
pulsate in the fundamental mode like the first two types of pulsators,
whereas their effective temperature is similar to the third type.
The core helium-burning phase in hot stars may take place only if they
suffer significant mass loss (about 75 \%). This emphasises the connection
to sdOB stars. However, although BLAPs and hot subdwarfs have comparable
effective temperatures, as seen in Fig.~\ref{fig:hr}, the two classes
of hot stars differ significantly in luminosity, by an order of magnitude.
Luminosity is inversely proportional to the surface gravity, and thus BLAPs
have much lower gravity (by an order of magnitude) than hot subdwarfs.
Relatively low gravity and high-amplitude pulsations
point to the presence of inflated envelopes in the new pulsators.

\begin{figure*}[h!]
\begin{center}
\includegraphics[width=0.9\textwidth]{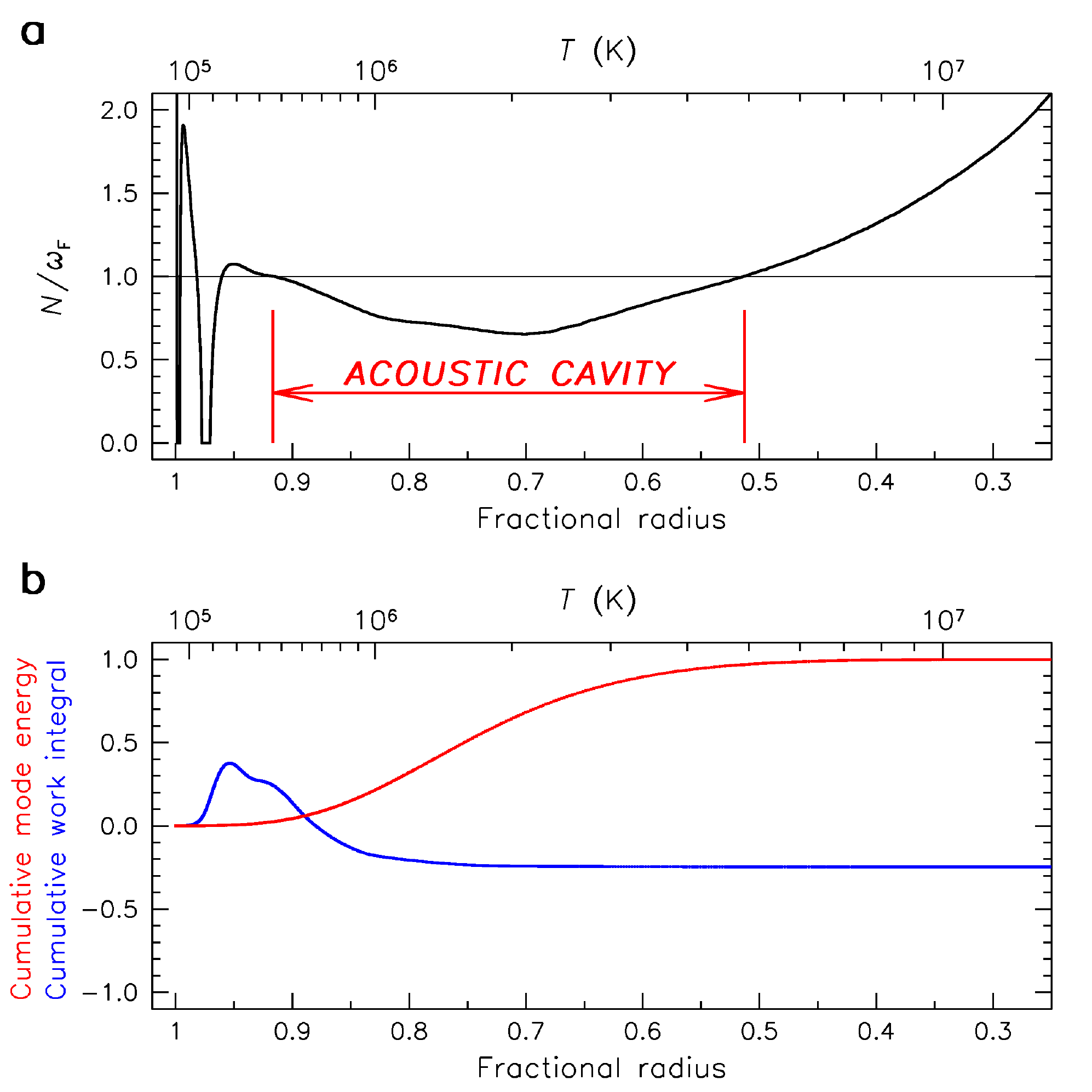}
\end{center}
\caption{\textbf {Proposed envelope model for the new pulsators.}
\textbf {a}, Pulsations are confined to an extended acoustic cavity
in the star envelope where the ratio of Brunt-V\"ais\"al\"a frequency
to the fundamental-mode frequency $N/\omega_{\rm F}<1$.
The bottom of the envelope is located very deep, at about 0.25 of the star
radius from its center or at a temperature of $T\approx2\times10^7$~K. Hence,
the star possesses a giant-like structure. In this sense, it is similar
to Cepheids and RR Lyrae-type stars in which low-order radial modes are preferred.
\textbf {b}, Nearly total contribution to the mode energy (red line)
comes from the acoustic cavity. The driving effect arises in a
narrow evanescent layer above the acoustic cavity around the maximum
of the iron opacity at $T\approx2\times10^5$~K. The damping takes
place close to the surface of the star where the work declines (blue line).
At the bottom of the envelope, the work drops below zero, implying
mode stability. Our simplistic model ignores the iron levitation.}
\label{fig:model}
\end{figure*}

To explore the idea of an inflated envelope, we have calculated a uniform
envelope model adopting mean values of the parameters derived from our
observations of the prototype object OGLE-BLAP-001 (Fig.~\ref{fig:model}).
The observed helium-to-hydrogen ratio of ${\rm log}~N({\rm He})/N({\rm H})=-0.55$
dex translates to the hydrogen and helium mass fractions of $X=0.46$ and
$Y=0.52$, respectively, assuming the solar metallicity of $Z=0.02$.
The model is calculated downward from the surface of the star to a
temperature of $T=2\times10^7$~K, low enough to neglect
the hydrogen burning. Our model shows that a significant
driving in the region of the metal opacity bump around
$T\approx2\times10^5$~K indeed occurs at the observed period of
$P=28.26$~min and surface temperature of $T_{\rm eff}=30,800$~K at star
luminosity of about ${\rm log}~L/L_{\odot}\approx2.6$. According to stellar
evolutionary models \cite{2006ApJ...642..797P}, such luminosity is
generated within the helium-burning core of a mass of about 1.0~$M_{\odot}$.
It can be formed in the evolution of a star with the
zero-age main sequence mass $M_{\rm ZAMS}\approx5$~M$_{\odot}$.
Pulsations are confined to an extended acoustic cavity in the star
envelope whose mass is only 2.5\% of the total mass
(Fig.~\ref{fig:model}). The excitation takes place in a narrow layer
above the acoustic cavity around the Z-bump. This bump is not only
responsible for driving pulsations in massive main-sequence B stars
($\beta$~Cep and SPB type), but also in evolved stars such as sdOBs on
the extreme horizontal branch. The cumulative work integral ($W$) at
the bottom of the envelope is below zero, implying mode stability.
This is expected because our model does not account for the radiative
levitation of iron, which plays an important role in exciting pulsation
modes in sdOB stars \cite{1997ApJ...483L.123C}. Importantly,
the first overtone is even more stable, as $W$ is lower than
for the fundamental mode.

A lower mass loss is required for the second possibility: that is,
BLAPs are stripped red giants with luminosities of about
${\rm log}~L/L_{\odot}\approx2.3$. In this case, the energy is
produced in the hydrogen-burning shell above the degenerate helium core
of a mass of $\sim0.3$~M$_{\odot}$. Such a configuration takes place
in stars of $M_{\rm ZAMS}\approx1$~M$_{\odot}$ before helium flash.
The second model seems more likely not only because the requirements
on mass loss are less severe, but also because it better reproduces
the observed gravity (Table~2). Furthermore, we are closer to the
fundamental mode instability. Gravities determined from the spectra
for the three other pulsators are slightly lower than for the prototype
object, but this is in agreement with their longer pulsation periods.

We may conclude that the pulsation properties of BLAPs are
remarkably similar to those of classical Cepheids and RR Lyrae-type
stars. However, there is a connection to sdOBs, as our stars
must have lost a significant fraction of the ZAMS mass to reach
higher surface temperature and gravity, as measured.
In the Hertzsprung-Russell diagram, the newly discovered pulsators
occupy a space where no pulsating variables had previously been known
(Fig.~\ref{fig:hr}). In terms of driving properties, BLAPs
are presumably related to the narrow instability domain recently
found at somewhat lower temperatures \cite{2016MNRAS.458.1352J}.
More accurate measurements of the gravity would allow for an independent
estimate of mass and luminosity, thus distinguishing between the two
proposed models for the new pulsators. The derived luminosities
of ${\rm log}~L/L_{\odot}=2.3$ and 2.6 for the prototype object
OGLE-BLAP-001 translate to its absolute brightness $M_V$ of +1.95
in the less massive case and +1.20~mag in the the more massive case.
Using the most recent extinction map \cite{2013ApJ...769...88N},
we find that all variables observed toward the Galactic bulge, except
for the brightest object OGLE-BLAP-009, are located at distances between
6 and 12~kpc, that is, within the bulge. Assuming an intrinsic colour
of the new pulsators of $(V-I)_0=-0.29$~mag and the standard interstellar
extinction law ($R_{V,VI}=2.46$), we obtain the distance to the prototype
variable of about 5.9 and 8.3~kpc, for the less massive
case and the more massive one, respectively. OGLE-BLAP-009 is the
closest of all pulsators. Like the prototype, it is located in the
Galactic Disk. The calculated distance is either of about 2.9
or 4.1~kpc, for the less massive case and more massive one, respectively,
assuming a higher luminosity of this object by 0.2~dex
than in the prototype and non-standard extinction law ($R_{V,VI}=2.14$)
towards the Galactic Bulge \cite{2003ApJ...590..284U}.

The formation scenario of BLAPs remains a mystery. Rejection of a significant
fraction of the stellar mass in the evolution of a single isolated low-mass
star is impossible. Such a huge mass loss could take place during an encounter
of the star with the central supermassive black hole. This has been
proposed to solve the problem of the absence of luminous red giants
in the inner parts of the Galactic Bulge \cite{2016ApJ...823..155K}.
Our variables, however, are not located close to the Galactic Centre.
They seem to be distributed over the whole bulge and two of them
reside in the Galactic Disk. Lines in the obtained spectra are
almost unshifted with respect to their laboratory positions,
indicating low radial velocities of BLAPs. This does not support
a runaway scenario resulting from such an encounter. A more
realistic explanation for the origin of our objects seems to be binary
evolution through mass transfer and common envelope ejection. This
is a widely accepted formation channel of sdB stars, as most of them
are observed in binary systems \cite{2002MNRAS.336..449H}. It is supposed
that single sdB stars are remnants of mergers between two helium white dwarfs
\cite{2016MNRAS.463.2756H}. In the case of BLAPs, we cannot exclude
the possibility that some of them might be in binary systems with
a companion faint enough to be detected in our data. The very small number
of BLAPs in comparison, for instance, to several hundreds of known sdB
stars, points to a rare episode in the stellar evolution.

\clearpage

\begin{methods}

\subsubsection*{Photometric observations and reductions}

The OGLE project operates the 1.3-m Warsaw Telescope located at Las Campanas
Observatory, Chile. Since the installation of a 32-detector mosaic CCD
camera with 1.4~deg$^2$ field of view in 2010, the project has been in its
fourth phase (OGLE-IV \cite{2015AcA....65....1U}). The previous phase
(OGLE-III \cite{2008AcA....58...69U}) was conducted in the years
2001--2009. Observations are taken mainly in the $I$ filter, whereas the
$V$ filter is used to secure colour information. The number of
$I$-band measurements collected by OGLE-IV towards the Galactic Bulge
in the years 2010--2016 varies greatly between individual fields,
from about 100 observations in the least-sampled fields to over
14,000 observations in the most frequently monitored areas.
Field cadence of the $I$-band data is from 20~min (or about 30 epochs
per night) up to 2 days. The total number of collected $V$-band measurements
is much smaller, between 8 and 140 observations. Reduction of the OGLE images
is performed with the help of difference image analysis technique
\cite{2000AcA....50..421W}, which provides very accurate photometry
in dense stellar fields (3\% for 18~mag, 10\% for 19.5~mag). The initial
search for periodic variability in nearly 400 million $I$-band light
curves from the Galactic Bulge was done using the publicly available FNPEAKS
code for periods longer than 0.01~day. The blue large-amplitude pulsators
presented here were serendipitously found during a visual inspection of
light curves with a signal-to-noise ratio higher than 5 and periods below 0.22~days
\cite{2015AcA....65...39S}. The TATRY code \cite{1996ApJ...460L.107S}
was used for precise determination of the periods after converting the
moments of observations from heliocentric Julian date (HJD) to the more accurate
barycentric Julian date (BJD$_{\rm TDB}$). We calculated the rate of the
period change for objects observed both in OGLE-III and OGLE-IV as
$$
r=\frac{\Delta P}{\Delta t}\frac{1}{P_{\rm IV}}
=\frac{P_{\rm IV}-P_{\rm III}}{t_{\rm IV}-t_{\rm III}}\frac{1}{P_{\rm IV}}.
$$
This simple method was used because of very small period changes observed
in our new pulsators. We found that dividing the data into separate seasons
did not lead to conclusive results.

An independent photometric campaign for the prototype object OGLE-BLAP-001,
located in the Galactic Disk, was conducted in 2016. The campaign helped us
to estimate the period change in this object after 10 years. It also
allowed us to determine the phase coverage of our follow-up spectroscopic
observations. The photometric data were obtained with the Direct CCD Camera
on the 1.0-m Swope telescope at Las Campanas Observatory on the following
four nights in 2016: 30--31 January, 2 February, and 26 April. A total of
229 $I$-band images with an exposure time of 120~s and 62 $V$-band images
with exposure time 180~s were collected. All images were de-biased and flat-fielded
using the IRAF package. The photometry was extracted with the help of difference
image analysis. A reference frame was constructed by combining 11 individual
images with seeing $\leq1.5$\arcsec. Profile photometry for the reference frame
was extracted with the DAOPHOT/ALLSTAR package \cite{1987PASP...99..191S}.
We used these measurements to transform the light curve from differential
flux units into magnitudes.

\subsubsection*{Spectroscopic observations and reductions}

The pair of spectra of the prototype object OGLE-BLAP-001 was obtained
with the 6.5-m Magellan-Baade telescope at the Las Campanas Observatory
on 22 April 2016. We used the Magellan Echellette (MagE) spectrograph
with a slit 10\arcsec long and 1.0\arcsec wide, giving an average
spectral resolution of $R\approx4100$ and covering wavelengths
between 3,200 and 10,000~\angstrom. Each spectrum was exposed for 900~s with
a read-out time of 30~s. Based on photometry from the Swope telescope,
we conclude that the spectra were taken in the following phase ranges:
0.162--0.693 and 0.711--1.242. Initial reductions (that is, flat-field correction,
rectification of spectral orders and wavelength calibration) were done
with the CarPy software on the site. In the next step, the orders were
normalized and combined using utilities provided in the IRAF package.

The low-resolution spectroscopic follow-up of objects OGLE-BLAP-009,
OGLE-BLAP-011, and OGLE-BLAP-014 was performed at the 8.1-m Gemini-South
telescope with the Gemini Multi-Object Spectrograph (GMOS) under programme
GS-2016A-Q-71 (Principal Investigator R. Angeloni). The observations
were conducted in regular queue mode between March and June 2016 on gray
nights ($V$-band sky brightness $>19.5$ mag~arcsec$^{-2}$), no
photometric conditions (maximum extinction of 0.3~mag over the nominal
atmospheric one) and seeing $\lesssim1.1$\arcsec. GMOS was configured
with the 1.0\arcsec slit and the B600\_G5323 grating centered at 5,500~\angstrom,
capable of delivering a resolution of $R\approx800$ over a wavelength
range of 4,000--7,000~\angstrom. Four 300~s exposures were taken for each
target following an optimized spatial and spectral dithering pattern:
it aimed at filling the small gaps in both the spatial and spectral
coverage due to the slit bridges (necessary in order to keep the
330\arcsec-long slit stable) and due to the gaps between the detectors,
respectively. The data frames were finally processed with the Gemini
IRAF package (v1.13.1 under Ureka v1.5.1) following the standard procedures
for long-slit data reduction.

The spectra were fitted with line-blanketed non-local thermodynamic
equilibrium (non-LTE) model atmospheres computed using the public codes TLUSTY
and SYNSPEC assuming a plane-parallel geometry \cite{1995ApJ...439..905L}.
The model grid covers the following parameter ranges:
20,000~K \textless\ $T_{\rm eff}$ \textless\ 50,000~K,
4.6 \textless\ log~$g$ \textless\ 6.4, and
$-4.0$ \textless\ log~$N$({\rm He})/$N$({\rm H}) \textless\ 0.0.
The metallic elements included in the models follow a typical abundance
for hot subdwarfs \cite{2008ApJ...678.1329B}: solar abundances for N, S
and Fe, and one tenth solar for C, O and Si.
The best-fit parameters are obtained via a $\chi^2$ minimization procedure
that relies on the method of Levenberg-Marquardt, based on a steepest-descent
method \cite{1994ApJ...432..351S}. Normalised hydrogen as well as helium
lines of both the observed and model spectra (previously convolved with
a Gaussian matching the observed resolution) are thus compared.

The same procedure was used for fitting the Gemini-South spectra of the
three follow-up objects; however, for these stars our initial fits indicated
surface gravities outside the lower limit of our grid (${\rm log}~g<4.6$).
To avoid extrapolation, we recomputed an extended grid,
this time covering the range $3.8<{\rm log}~g<5.6$. These new models
are computed in non-LTE but only include opacity from hydrogen and
helium, because computing a grid of fully line-blanketed models requires
a considerable time. We expect the effect of metallicity
on the derived parameters to be rather small for effective temperatures
below 35,000~K \cite{2014ApJ...788...65L}, and, given the modest signal-to-noise
ratio of the spectra, the formal fitting uncertainties are much larger
than any systematic that could be caused by the metallicity of the models.

\subsubsection*{Model of the envelope}

To investigate radial-mode stability in the new pulsators, we applied
a linear non-adiabatic equilibrium envelope model \cite{1969AcA....19....1P}
updated for the use of modern atomic data (OPAL opacities
\cite{1996ApJ...464..943I}). A standard ratio of mixing length to
pressure scale-height $\alpha=1.5$ was adopted. The effect of rotation
was ignored. The envelope model was calculated for various star
masses assuming radial F-mode pulsations with $P_{\rm F}=0.01962$~d,
effective temperature of 30,800~K (${\rm log}~T_{\rm eff}=4.48855$)
and uniform chemical composition with abundances derived for the
prototype object OGLE-BLAP-001. Pulsations are regarded as a small
departure from the dynamical and thermal equilibrium
\cite{1977AcA....27...95D}. The period is insensitive to the interior
as long as the Brunt-V\"ais\"al\"a frequency at the envelope base, $N$,
exceeds the mode frequency, $\omega_{\rm F}=2\pi/P_{\rm F}$.

\end{methods}

\clearpage


\begin{addendum}
\item[Correspondence] Correspondence and requests for materials should be
addressed to P.P.\\(pietruk@astrouw.edu.pl).

\item[Acknowledgements]
We thank M. Kubiak and G. Pietrzy\'nski, former
members of the OGLE team, for their contribution to the collection
of the OGLE photometric data over the past years. The OGLE project
has received funding from the National Science Centre,
Poland (grant number MAESTRO 2014/14/A/ST9/00121 to A.U.).
M.L. acknowledges support from the Alexander von Humboldt Foundation.
The Las Campanas Observatory, which hosts the Warsaw Telescope,
Swope Telescope and Magellan Telescopes is operated by the
Carnegie Institution for Science. The Gemini Observatory is operated
by the Association of Universities for Research in Astronomy, Inc.,
under a cooperative agreement with the NSF on behalf of the Gemini
partnership: the National Science Foundation (United States),
the National Research Council (Canada), CONICYT (Chile), Ministerio de Ciencia,
Tecnolog\'ia e Innovaci\'on Productiva (Argentina), and Minist\'erio
da Ci\^encia, Tecnologia, Inova\c{c}\~oes e Comunica\c{c}\~oes (Brazil).

\item[Author contributions]
P.P. coordinated the research, obtained and analysed part of the
observations and prepared the manuscript.
W.A.D. proposed the envelope model and calculated its characteristics.
M.L. fitted model atmospheres to the spectroscopic data.
R.A. obtained and reduced Gemini spectra.
R.P. and F.diM. obtained part of the follow-up photometric observations.
The remaining authors, including also P.P. and R.P., collected the
OGLE observations. All authors commented on the manuscript and were
involved in the science discussion.

\item [Data Availability Statement]
The data that support the plots within this paper and other findings of this
study are available from the corresponding author upon reasonable request.
The time-series photometry of the new variables is available to the
astronomical community from the OGLE Internet Archive at\\
ftp://ftp.astrouw.edu.pl/ogle/ogle4/OCVS/BLAP/\\

\item [Competing interests]
The authors declare no competing financial interests.

\end{addendum}


\newpage

\newpage

\newpage

\newpage

\newpage

\newpage

\newpage


\newpage
\begin{table}[h!]
\caption{\small Observational parameters of the discovered BLAPs:
equatorial coordinates, mean brightness and amplitude in the $V$
and $I$ bands, pulsation period, and rate of the period change. The rate and
its uncertainty is calculated based on two epochs representing the third and
the fourth (current) phase of the OGLE survey from years 2001--2009 and 2010--2016,
respectively.}
\begin{center}
{\scriptsize
\begin{tabular}{lllllllll}
\hline
\hline
Variable name & RA(2000.0)  & Dec(2000.0) & $\langle V \rangle$ & $\langle I \rangle$  & $A_V$ & $A_I$ & $P$ & $r$ \\
 & ($^{\rm h}$:$^{\rm m}$:$^{\rm s}$) & (\degr:\arcmin:\arcsec) & (mag) & (mag) & (mag) & (mag) & (min) & ($\times 10^{-7}$ y$^{-1}$) \\
\hline
OGLE-BLAP-001 & 10:41:48.77 & $-61$:25:08.5 & 17.706 & 17.223 & 0.411 & 0.236 & 28.26 & $+2.9 \pm3.7$ \\
OGLE-BLAP-002 & 17:43:58.02 & $-19$:16:54.1 & 18.892 & 17.953 & 0.317 & 0.357 & 23.29 & \\
OGLE-BLAP-003 & 17:44:51.48 & $-24$:10:04.0 & 19.195 & 18.103 & 0.273 & 0.229 & 28.46 & $+0.82 \pm0.32$ \\
OGLE-BLAP-004 & 17:51:04.72 & $-22$:09:03.4 & 18.813 & 17.528 & 0.426 & 0.261 & 22.36 & \\
OGLE-BLAP-005 & 17:52:18.73 & $-31$:56:35.0 & 19.989 & 18.767 & 0.333 & 0.298 & 27.25 & $+0.63 \pm0.26$ \\
OGLE-BLAP-006 & 17:55:02.88 & $-29$:50:37.5 & 18.354 & 17.384 & 0.279 & 0.231 & 38.02 & $-2.85 \pm0.31$ \\
OGLE-BLAP-007 & 17:55:57.52 & $-28$:52:11.0 & 19.268 & 18.456 & 0.330 & 0.282 & 35.18 & $-2.40 \pm0.51$ \\
OGLE-BLAP-008 & 17:56:48.26 & $-32$:21:35.6 & 18.732 & 17.590 & 0.284 & 0.194 & 34.48 & $+2.11 \pm0.27$ \\
OGLE-BLAP-009 & 17:58:48.20 & $-27$:16:53.7 & 15.650 & 15.071 & 0.286 & 0.241 & 31.94 & $+1.63 \pm0.08$ \\
OGLE-BLAP-010 & 17:58:59.22 & $-35$:18:07.0 & 17.231 & 16.919 & 0.383 & 0.344 & 32.13 & $+0.44 \pm0.21$ \\
OGLE-BLAP-011 & 18:00:23.24 & $-35$:58:03.1 & 17.254 & 16.977 & 0.222 & 0.286 & 34.87 & \\
OGLE-BLAP-012 & 18:05:44.20 & $-30$:11:15.2 & 18.263 & 17.622 & 0.350 & 0.343 & 30.90 & $+0.03 \pm0.15$ \\
OGLE-BLAP-013 & 18:05:52.70 & $-26$:48:18.0 & 19.244 & 18.190 & 0.249 & 0.221 & 39.33 & $+7.65 \pm0.67$ \\
OGLE-BLAP-014 & 18:12:41.79 & $-31$:12:07.8 & 16.793 & 16.680 & 0.339 & 0.264 & 33.62 & $+4.82 \pm0.39$ \\
\hline
\end{tabular}}
\end{center}
\end{table}

\newpage
\begin{table}[h!]
\caption{\small Derived parameters of the envelope model for various star masses.
The bottom of the envelope is placed at the radius $r=r_{\rm core}$ where
temperature is $2\times10^7$~K. Normalized driving rate $\eta$ is calculated
for the fundamental mode (F) and first overtone (1O). The two versions of
evolutionary models consistent with similar values of $L(M_{\rm core})$
are given in the last column. Heavy $\sim1.0$~M$_{\odot}$ helium cores
are formed in the evolution of stars with $M_{\rm ZAMS}\approx5$~M$_{\odot}$.
In this case, a huge mass loss is required to create a BLAP. A much lower mass loss
is needed if the BLAPs are shell hydrogen-burning objects with degenerate helium
cores. At $M=0.3$~M$_{\odot}$, the core has similar mass and luminosity as a
$M_{\rm ZAMS}\approx1$~M$_{\odot}$ star on the red giant branch well before
helium flash.}
\begin{center}
{\scriptsize
\begin{tabular}{lllllllll}
\hline
\hline
$M/M_{\odot}$ & ${\rm log}~T_{\rm eff}$ & ${\rm log}~L/L_{\odot}$ & ${\rm log}~g$ & $M_{\rm core}/M_{\odot}$ & $r_{\rm core}/R$ & $\eta_{\rm F}$ & $\eta_{\rm 1O}$ & Model \\
\hline
0.30 & 4.4886 & 2.2572 & 4.5496 & 0.2983 & 0.1061 & $-0.239$ & $-0.473$ & shell H-burning \\
0.35 & 4.4886 & 2.3039 & 4.5710 & 0.3474 & 0.1161 & $-0.230$ & $-0.417$ & \\
0.40 & 4.4886 & 2.3440 & 4.5901 & 0.3963 & 0.1255 & $-0.225$ & $-0.380$ & \\
0.45 & 4.4886 & 2.3787 & 4.6073 & 0.4450 & 0.1343 & $-0.223$ & $-0.354$ & \\
0.50 & 4.4886 & 2.4097 & 4.6229 & 0.4934 & 0.1427 & $-0.222$ & $-0.334$ & \\
0.60 & 4.4886 & 2.4626 & 4.6500 & 0.5893 & 0.1582 & $-0.224$ & $-0.306$ & \\
0.80 & 4.4886 & 2.5466 & 4.6926 & 0.7777 & 0.1852 & $-0.234$ & $-0.275$ & \\
1.00 & 4.4886 & 2.6123 & 4.7250 & 0.9613 & 0.2083 & $-0.247$ & $-0.261$ & core He-burning \\
1.20 & 4.4886 & 2.6659 & 4.7509 & 1.1399 & 0.2286 & $-0.258$ & $-0.254$ & \\
\hline
\end{tabular}}
\end{center}
\end{table}

\end{document}